\documentclass[aps,preprint,nofootinbib,superscriptaddress]{revtex4}%
\usepackage[english]{babel}
\usepackage{amsmath, amssymb, mathrsfs, graphics, setspace, graphicx,
	epstopdf, cancel, color}%
\usepackage{amsmath}%
\usepackage{amsfonts}%
\usepackage{amssymb}%
\usepackage{graphicx}
\providecommand{\U}[1]{\protect\rule{.1in}{.1in}}
\numberwithin{equation}{section}
\newcommand{\nn}{\nonumber}

\newcommand{\bnabla}{\bar{\nabla}}
\newcommand{\tnabla}{\tilde{\nabla}}
\newcommand{\hpi}{\hat{\pi}}
\begin{document}
	\title{Fluid/Gravity Correspondence with Scalar Field and Electromagnetic Field}
	
	\author{Chia-Jui Chou}
	\email{agoodmanjerry.ep02g@nctu.edu.tw}
	\affiliation{Department of Electrophysics, National Chiao Tung University, Hsinchu, ROC}
	\author{Xiaoning Wu}
	\email{wuxn@amss.ac.cn}
	\affiliation{Institute of Mathematics, Academy of Mathematics and System Science, CAS, China}
	\author{Yi Yang}
	\email{yiyang@mail.nctu.edu.tw}
	\affiliation{Department of Electrophysics, National Chiao Tung University, Hsinchu, ROC}
	\author{Pei-Hung Yuan}
	\email{phyuan.py00g@nctu.edu.tw}
	\affiliation{Institute of Physics, National Chiao Tung University, Hsinchu, ROC}
	
	\begin{abstract}
		We consider fluid/gravity correspondence in a general rotating black hole background with scalar and electromagnetic fields. Using the method of Petrov-like boundary condition, we show that the scalar and the electromagnetic fields contribute external forces to the dual Navier-Stokes equation and the rotation of black hole induces the Coriolis force.
	\end{abstract}
	\maketitle
	\tableofcontents
	\newpage
	
	%TCIMACRO{\TeXButton{equation number}{\setcounter{equation}{0}
	%\renewcommand{\theequation}{\arabic{section}.\arabic{equation}}}}%
	%BeginExpansion
	\setcounter{equation}{0}
	\renewcommand{\theequation}{\arabic{section}.\arabic{equation}}%
	%EndExpansion

	\section{Introduction}
	
	Fluid/gravity correspondence is one of the most interesting topics arose from studying the surface effect of black hole horizon\cite{Damour1, damour1978, Damoursurface, 0802.4169}. It gives an additional support to AdS/CFT correspondence. After the pioneer works by Policastro, Starinets and Son \cite{0104066, 0205052, 0309213, 0704.0240}, Bhattacharrya et al. constructed the correspondence involving nonlinear effects of fluid \cite{0712.2456, 0810.1545, 1511.01377} and
	Eling, Oz et al. established the way of studying fluid/gravity correspondence by membrane paradigm \cite{eling1, eling2,eling3, 0809.4512}.
	
    Based on the AdS/CFT correspondence, the variation of the boundary radius $r$ corresponds to the RG flow of the dual field theory. In order to consider the behavior of RG flow, Strominger et al. studied the fluid/gravity correspondence on a finite cutoff of space-time \cite{1006.1902, 1101.2451}. In \cite{1104.5502}, they developed a new method to consider the fluid/gravity correspondence on horizon.
	The Petrov-like boundary condition provides enough constraints to reduce the degrees of freedom of the gravitational perturbations to the degrees of freedom of a fluid on $\Sigma_{c}$.
	By properly identifying the velocity and pressure of the fluid near horizon to the corresponding perturbation terms in the Brown-York tensor, one is able to obtain the Navier-Stokes equation on $\Sigma_c$ from Gauss-Codazzi equation.
	
    The method of using Petrov-like condition on a cutoff surface near horizon is a very powerful way to study fluid/gravity correspondence. By using this method, one is able to obtain the Navier-Stokes equation explicitly from Einstein equation in various space-time backgrounds. For example, fluid/gravity correspondence with Chern-Simons term and Gauss-Bonnet term were considered in \cite{1408.6488}, space-time coupled with scalar fields and electromagnetic fields were considered in \cite{1401.6487, 1511.08281} and \cite{1204.0959, 1310.4181}, and Einstein-Dilaton-Axion theory was considered in \cite{1508.04972}.
	
	All the works mentioned above studied fluid/gravity correspondence under some specific space-time backgrounds by choosing certain concrete metrics.
	In \cite{1303.3736}, general non-rotating black holes in vacuum background have been studied using initial value formulation without assuming any specific metric and Navier-Stokes equation of the incompressible fluid was obtained.
	Recently, general rotating black holes in vacuum background was also studied using similar method and an interesting effect was found that the angular momentum of black hole causes a Coriolis force term in the dual fluid equation \cite{1511.08691}.
	
	In this paper, we would like to study the effects caused by both matter fields and black hole rotation in fluid/gravity correspondence. One of the motivations to do this is that, based on the basic idea of the previous works on holographic QCD \cite{1012.1864,1108.2029,1209.4512,1301.0385,1406.1865,1506.05930} and holographic condensed matter theories \cite{0809.4870,1005.4690, 1101.3332}, the Einstein-Maxwell-scalar system in the bulk has been successfully used to study the dual field theories at the boundary. Such models have been proved to be very useful in gauge/gravity correspondence. The explicit question we want to ask here is:  how will Navier-Stokes equation be modified in general by those matter fields? For this purpose, we will consider an Einstein-Maxwell-scalar system near an isolated horizon.
	We find that the matter fields of scalar and electromagnetic fields only contribute external force terms to the Navier-Stokes equation.
	
	The paper is organized as follows:
	In section \ref{II}, we introduce the action and equations of motion of the Einstein-Maxwell-scalar system and express the components of the null tetrad and metric of $(p+2)$-dimensional space-time by their initial data on the horizon. 	In section \ref{III}, we introduce a $(p+1)$-dimensional cutoff hypersurface $\Sigma_c$ near the horizon and calculate the constraints given by Petrov-like condition $C_{lilj} = 0$ on $\Sigma_c$ in near-horizon and non-relativistic limit.
	We then apply the Petrov-like condition to Gauss-Codazzi equation to get the Navier-Stokes equation in section \ref{IV}.
	Section \ref{V} includes conclusion and discussion on our results comparing with other related works.
	
	%TCIMACRO{\TeXButton{equation number}{\setcounter{equation}{0}
	%\renewcommand{\theequation}{\arabic{section}.\arabic{equation}}}}%
	%BeginExpansion
	\setcounter{equation}{0}
	\renewcommand{\theequation}{\arabic{section}.\arabic{equation}}%
	%EndExpansion

	\section{$p+2$-dimensional Space-time Setup}
	\label{II}
	
	\subsection{Einstein-Maxwell-scalar System}
	
	We consider a $(p+2)$-dimensional space-time background with a complex scalar field coupled to an electromagnetic field. The action is
	\begin{equation} \label{action}
	S=\int d^{p+2}x \sqrt{-g}\left[ R - 2\Lambda - \frac{1}{2}%
	|D\phi|^2 - \frac{1}{2}V(\phi,\phi^*)-\frac{1}{4}f(\phi,\phi^*)F^{ab}F_{ab}\right],
	\end{equation}
	where $F_{ab}$ is the field strength of the Maxwell field $A_a$, $f(\phi, \phi^*)$ is the gauge kinetic function and $V(\phi, \phi^*)$ is the potential of the scalar field $\phi$. The field strength $F_{ab}$ and covariant derivative $D_a$ are defined as
	\begin{subequations}
		\begin{align}
		F_{ab} &  =\partial_{a}A_{b}-\partial_{b}A_{a},\\
		D_a &  =\nabla_{a}-ieA_{a},
		\end{align}
	\end{subequations}
	with the complex scalar field $\phi$ carrying charge ($-e$).
	In this paper, we will use different notations for different kinds covariant derivatives : $\nabla$ denotes the covariant derivative in $(p+2)$-dimensional space-time, $\bnabla$ denotes the covariant derivative on the $(p+1)$-dimensional time-like boundary $\Sigma_c$ and $\tnabla$ denotes the covariant derivative on the $p$-dimensional section $S^p$.
	
	Varying the Einstein-Maxwell-scalar action (\ref{action}) with the fields of $\phi$, $\phi^*$, $A^{a}$, and the metric $g^{ab}$, respectively, we obtain the equations of motion of each fields as
	\begin{subequations} \label{EOM}%
		\begin{align}
		0= &  D^2\phi - \frac{\partial V}{\partial\phi^*} - \frac{1}{2}\frac{\partial f}{\partial\phi^*}F^2 = D^2\phi^* - \frac{\partial V}{\partial\phi} - \frac{1}{2}\frac{\partial f}{\partial\phi}F^2,\\
		0= &  D^{b}(fF_{ab}) + ie\Big( \phi^*D_a\phi - \phi D_a^*\phi^* \Big),\\
		0= &  R_{ab} - \left( \frac{1}{2}R - \Lambda \right)g_{ab} - \frac{1}{2}(D_a\phi)^* (D_b\phi) - \frac{f}{2}F_{ac}F_b^{~c} + \frac{1}{4}g_{ab}\left( |D\phi|^2 + V + \frac{f}{2}F^2 \right).
		\end{align}
	\end{subequations}
	The Ricci scalar and Ricci tensor become
	\begin{equation}
	R=\frac{2(p+2)}{p}\Lambda + \frac{(p+2)}{2p}V + \frac{1}{2}|D\phi|^{2}+\frac{(p-2)}{4p}fF^{2},
	\end{equation}
	and
	\begin{equation}
	R_{ab}=\Lambda^{\prime}g_{ab}+J_{ab},\label{Ricci0}%
	\end{equation}
	where we have defined
	\begin{equation}
	\Lambda^{\prime}=\frac{8\Lambda + 2V - fF^2}{4p},\quad J_{ab} = \frac{1}{2}(D_{a}\phi)^*(D_{b}\phi) + \frac{f}{2}F_{ac}F_{b}^{~c}.\label{Jab}%
	\end{equation}
	We can solve the equations of motion order by order by expanding the fields $\phi$ and $A_{a}$ with small $r$,
	\begin{subequations}
		\begin{align}
		\phi(t,r,x^{i}) &  =\phi_{(0)}(t,x^{i})+\sum_{k=1}\frac{1}{k!}\phi
		_{(k)}(t,x^{i})r^{k},\\
		A_{a}(t,r,x^{i}) &  =A_{a(0)}(t,x^{i})+\sum_{k=1}\frac{1}{k!}A_{a(k)}%
		(t,x^{i})r^{k},
		\end{align}
		with $\phi_{(0)}(t,x^{i})$ and $A_{a(0)}(t,x^{i})$ are the initial data at $r=0$. We can use equations of motion (\ref{EOM}) to express the coefficients of the higher order terms $\phi_{(k)}$ and $A_{a(k)}$ in terms of the initial data $\phi_{(0)}(t,x^{i})$ and $A_{a(0)}(t,x^{i})$, although these higher order terms will not contribute to the Navier-Stokes equation.
		
		%TCIMACRO{\TeXButton{equation number}{\setcounter{equation}{0}
		%\renewcommand{\theequation}{\arabic{section}.\arabic{equation}}}}%
		%BeginExpansion
		\setcounter{equation}{0}
		\renewcommand{\theequation}{\arabic{section}.\arabic{equation}}%
		%EndExpansion

		\subsection{Newman-Penrose Tetrad and Metric}
		
		The fluid/gravity correspondence of general black holes in a vacuum space-time background has been studied in \cite{1303.3736}.
		It was shown that, in near-horizon limit and non-relativistic limit, one is able to obtain Navier-Stokes equation describing an incompressible fluid on the hypersurface.
		Now we are going to study the Navier-Stokes equation in the $(p+2)$-dimensional space-time background introduced above.
		
		We choose the black hole horizon to be an ``Isolated Horizon (IH)". According to \cite{0111067}, the definition of an Isolated Horizon is :
		
		\textbf{Definition 1} \textit{ (Isolated Hoizon in $\mathit{(p+2)}$-dimensional space-time)\newline
		Let $(M,g)$ be a $(p+2)$-dimensional space-time. $\mathcal{H}$ is a $(p+1)$-dimensional null sub-manifold in $(M,g)$ and $l$ is the future-directed null normal of $\mathcal{H}$. $\mathcal{H}$ is called a isolated horizon if
			\newline(1)~ $\mathcal{H}$ is diffeomorphic to the product $S^p\times\Re$, where $S^p$ is a $p$-dimensional space-like manifold, and the fibers of the projection
			\begin{equation*}
			\Pi : S^p \times \Re \rightarrow S^p
			\end{equation*}
			are null curves in $\mathcal{H}$;
			\newline(2)~ the expansion of l vanishes everywhere on $\mathcal{H}$, i.e. $\nabla_al^a = 0$ on $\mathcal{H}$;
			\newline(3)~ the stress-energy tensor $T_{ab}$ is such that $-T_{ab}l^{~b}|_{\mathcal{H}}$ is future-directed causal for any future-directed null vector $l^a$ and all field equations and Einstein equation hold in a neighbourhood of $\mathcal{H}$;
			\newline(4)~ let $\mathcal{D}$ denote the induced connection on $\mathcal{H}%
			$, $[\mathcal{L}_{l},\mathcal{D}] \hat{=} 0$. }

		To describe the geometry of null hypersurface near the horizon, we introduce
		the Bondi-like coordinate system. We use the null tetrad of the Newman-Penrose
		formalism $\{n,l,E_{I}\}$ with $I=1,2,\cdots,p$ and the coordinates
		$(t,r,x^{i})$ with $i=1,2,\cdots,p$. The inner product of the null tetrad is
	\end{subequations}
	\begin{subequations}
		\begin{gather}
		\langle n,l\rangle=\langle l,n\rangle=1,\\
		\langle E_{I},E_{J}\rangle=\delta_{IJ},\\
		\text{Otherwise}=0.
		\end{gather}
		Follow the same Bondi gauge and the Bondi-like coordinates $(t,r,x^{i})$ near horizon used in \cite{1303.3736, 1511.08691}, we have
	\end{subequations}
	\begin{equation}
	\nabla_{n}(n,l,E_{I})=0,
	\end{equation}
	with $t$ is the parameter of the null generator $l^a$ of $\mathcal{H}$, $x^i$ are the coordinates of $S^p$ and $r$ is the affine parameter of $n^a$ which is the null tetrad orthogonal to $l^a$ and $E_I^a$.
	One can choose the null tetrad as,
	\begin{subequations}
		\begin{align}
		n^{a} &  =(\partial_{r})^{a},\\
		l^{a} &  =(\partial_{t})^{a}+U(\partial_{r})^{a}+X^{i}(\partial_{i})^{a},\\
		E_{I}^{a} &  =W_{I}(\partial_{r})^{a}+e_{I}^{i}(\partial_{i})^{a},
		\end{align}
		where, $(U,X^{i},W_{I},e_{I}^{i})$ are functions of $(t,r,x^{i})$ and $\hat
		{U}=\hat{X^{i}}=\hat{W_{I}}=0$ with\newline\textquotedblleft\ $\hat{}$ "
		denoting the initial value on the horizon $\mathcal{H}$.
		
		The metric is
	\end{subequations}
	\begin{equation}
	g^{ab} = n^{a} l^{b} + l^{a} n^{b} + E^{a}_{I} E^{b}_{I},
	\end{equation}
	gives
	\begin{equation}
	{g}^{\mu\nu} =
	\begin{pmatrix}
	0 & 1 & 0\\
	1 & 2U + W_{I}W_{I} & X^{i} + W_{I}e^{i}_{I}\\
	0 & X^{j} + W_{I}e^{j}_{I} & e^{i}_{I}e^{j}_{I}%
	\end{pmatrix}
	,
	\end{equation}
	and
	\begin{equation}
	{g}_{\mu\nu} =
	\begin{pmatrix}
	-g^{rr} + g_{ij}g^{ri}g^{rj} & 1 & -g_{ij}g^{ri}\\
	1 & 0 & 0\\
	-g_{ij}g^{rj} & 0 & g_{ij}%
	\end{pmatrix}
	.
	\end{equation}	
	We define the spin connections of Newman-Penrose formalism using the null
	tetrad,
	\begin{equation}%
	\begin{split}
	& \alpha_{I} = -\langle l,\nabla_{I}n\rangle, \quad\pi_{I} = \langle
	E_{I},\nabla_{l}n\rangle,\\
	& \epsilon= \langle n,\nabla_{l}l\rangle, \quad\theta^{\prime}_{JI} = \langle
	E_{J},\nabla_{I}n\rangle.
	\end{split}
	\end{equation}
	$\epsilon|_{\mathcal{H}}$ is the surface gravity of $\mathcal{H}$ and is a constant.
	By Bondi gauge, it is easy to see that,
	\begin{equation}
	\alpha_{I} = -\pi_{I},
	\end{equation}
	where $\pi_{I}$ is related to the angular momentum of the horizon.
	
	\subsection{Components of the Metric}
	
	To solve the evolution of the horizon, we need to know cert initial data including the spin
	coefficients on $\mathcal{H}$, the Weyl tensor on $\mathcal{H}$, and the
	metric of $p$-sphere $S^p$ on $\mathcal{H}$. Other physical quantities near
	horizon can be expressed by these initial data.
	
	First, we expand the metric in small $r$ near horizon and express
	the coefficients by the initial data on $\mathcal{H}$. The leading order metric in $r$ can calculated by using the first Cartan structure equation,
	\begin{align}
	\lbrack n,E_{I}]= &  \nabla_{n}E_{I}-\nabla_{I}n=-\nabla_{I}n\nonumber\\
	= &  -\langle n,\nabla_{I}n\rangle l^{a}-\langle l,\nabla_{I}n\rangle
	n^{a}-\langle E_{J},\nabla_{I}n\rangle{E_{J}}^{a}\nonumber\\
	= &  (\partial_{r}W_{I})(\partial_{r})^{a}+(\partial_{r}e_{I}^{i}%
	)(\partial_{i})^{a},
	\end{align}%
	and
	\begin{align}
	\lbrack n,l] &  =\nabla_{n}l-\nabla_{l}n=-\nabla_{l}n\nonumber\\
	&  =-\langle l,\nabla_{l}n\rangle n^{a}-\langle n,\nabla_{l}n\rangle
	l^{a}-\langle E_{I},\nabla_{l}n\rangle E_{I}^{a}\nonumber\\
	&  =(\partial_{r}U)(\partial_{r})^{a}+(\partial_{r}X^{i})(\partial_{i})^{a}.
	\end{align}
	Therefore, we have the first derivative of the coefficients of metric,
	\begin{subequations}
		\label{first Cartan}%
		\begin{align}
		\partial_{r}U &  =\epsilon-\pi_{I}W_{I},\\
		\partial_{r}X^{i} &  =-\pi_{I}e_{I}^{i},\\
		\partial_{r}W_{I} &  =\alpha_{I}-\theta_{JI}^{\prime}W_{J},\\
		\partial_{r}e_{I}^{i} &  =-\theta_{JI}^{\prime}e_{J}^{i}.
		\end{align}
		The second Cartan equation gives us,
	\end{subequations}
	\begin{subequations}
		\label{second Cartan}%
		\begin{align}
		\partial_{r}\epsilon &  =R_{nlnl}-\alpha_{I}\pi_{I},\\
		\partial_{r}\pi_{I} &  =R_{nlIn}-\pi_{J}\theta_{IJ}^{\prime},\\
		\partial_{r}\alpha_{I} &  =R_{nInl}-\alpha_{J}\theta_{JI}^{\prime},\\
		\partial_{r}\theta_{JI}^{\prime} &  =R_{nIJn}-\theta_{KI}^{\prime}\theta
		_{JK}^{\prime}.
		\end{align}
		Then, by taking the derivative with respect to $r$ of equation
		(\ref{first Cartan}) and using the equation (\ref{second Cartan}), we get,
	\end{subequations}
	\begin{subequations}
		\begin{align}
		\partial_{r}^{2}U &  =R_{nlnl}-\alpha_{I}\pi_{I}-(R_{nlIn}-\theta_{IJ}%
		^{\prime}\pi_{J})W_{I}-(\alpha_{I}-\theta_{JI}^{\prime}W_{J})\pi_{I},\\
		\partial_{r}^{2}W_{I} &  =R_{nInl}-\alpha_{J}\theta_{JI}^{\prime}%
		-(R_{nIJn}-\theta_{JK}^{\prime}\theta_{KI}^{\prime})W_{J}-(\alpha_{J}%
		-\theta_{KJ}^{\prime}W_{K})W_{J},\\
		\partial_{r}^{2}X^{i} &  =(R_{nInl}+\theta_{IJ}^{\prime}\pi_{J})e_{I}%
		^{i}+(\theta_{JI}^{\prime}e_{J}^{i})\pi_{I},\\
		\partial_{r}^{2}e_{I}^{i} &  =(R_{nInJ}+\theta_{JK}^{\prime}\theta
		_{KI}^{\prime})e_{J}^{i}+(\theta_{JI}^{\prime}e_{J}^{i})\theta_{JI}^{\prime}.
		\end{align}
		Using the above equations and the fact that $\pi_{I}=-\alpha_{I}$ and $\hat
		{U}=\hat{W}_{I}=\hat{X}^{i}=0$, we obtain,
	\end{subequations}
	\begin{subequations}
		\begin{align}
		U= &  \hat{\epsilon}r+\frac{1}{2}(\hat{R}_{nlnl}+2|\hat{\pi}|^{2}%
		)r^{2}+O(r^{3}),\\
		W_{I}= &  -\hat{\pi}_{I}r+\frac{1}{2}(\hat{R}_{nInl}+2\hat{\theta}%
		_{IJ}^{\prime}\hat{\pi}_{J})r^{2}+O(r^{3}),\\
		X^{i}= &  -\hat{\pi}_{I}\hat{e}_{I}^{i}r+\frac{1}{2}(\hat{R}_{nInl}\hat{e}%
		_{I}^{i}+2\hat{\theta}_{IJ}^{\prime}\hat{\pi}_{I}\hat{e}_{J}^{i})r^{2}%
		+O(r^{3}),\\
		e_{I}^{i}= &  \hat{e}_{I}^{i}-\hat{\theta}_{IJ}^{\prime}\hat{e}_{J}^{i}%
		r+\frac{1}{2}(\hat{R}_{nInJ}\hat{e}_{J}^{i}+2\hat{\theta}_{IJ}^{\prime}%
		\hat{\theta}_{JK}^{\prime}\hat{e}_{K}^{i})r^{2}+O(r^{3}),
		\end{align}
		which lead to the following metric,
	\end{subequations}
	\begin{subequations}
		\begin{align}
		g^{tr} &  =1,\\
		g^{rr} &  =2\hat{\epsilon}r+(\hat{R}_{nlnl}+3|\hat{\pi}|^{2})r^{2}%
		+O(r^{3}),\\
		g^{ri} &  =-2\hat{\pi}_{I}\hat{e}_{I}^{i}r+(\hat{R}_{nInl}\hat{e}_{I}%
		^{i}+3\hat{\theta}_{IJ}^{\prime}\hat{\pi}_{I}\hat{e}_{J}^{i})r^{2}+O(r^{3}),\\
		g^{ij} &  =\hat{e}_{I}^{i}\hat{e}_{I}^{j}-2\hat{\theta}_{IJ}^{\prime}\hat
		{e}_{I}^{i}\hat{e}_{J}^{j}r+(\hat{R}_{nInJ}\hat{e}_{I}^{i}\hat{e}_{J}%
		^{j}+3\hat{\theta}_{IK}^{\prime}\hat{\theta^{\prime}}_{JK}\hat{e}_{I}^{i}%
		\hat{e}_{J}^{j})r^{2}+O(r^{3}),
		\end{align}
		and
	\end{subequations}
	\begin{subequations}
		\begin{align}
		g_{tt} &  =-2\hat{\epsilon}r-(\hat{R}_{nlnl}-|\hat{\pi}|^{2})r^{2}%
		+O(r^{3}),\\
		g_{tr} &  =1,\\
		g_{ti} &  =2\hat{\pi}_{I}\hat{e}_{I}^{j}\hat{g}_{ij}r-(\hat{R}_{nInl}\hat
		{e}_{I}^{j}\hat{g}_{ij}-\hat{\theta}_{IJ}^{\prime}\hat{\pi}_{I}\hat{e}_{J}%
		^{j}\hat{g}_{ij})r^{2}+O(r^{3}),\\
		g_{ij} &  =\hat{g}_{ij}+2\hat{\theta}_{IJ}^{\prime}\hat{e}_{I}^{k}\hat{e}%
		_{J}^{m}\hat{g}_{ik}\hat{g}_{jm}r+O(r^{2}).
	\end{align}
	\end{subequations}
		
	%TCIMACRO{\TeXButton{equation number}{\setcounter{equation}{0}
	%\renewcommand{\theequation}{\arabic{section}.\arabic{equation}}}}%
	%BeginExpansion
	\setcounter{equation}{0}
	\renewcommand{\theequation}{\arabic{section}.\arabic{equation}}%
	%EndExpansion
	
	\section{$p+1$-dimensional Cutoff Hypersurface}
	\label{III}
	\subsection{Induced Metric and Extrinsic Curvature}
		Brown-York tensor on the hypersurface $\Sigma_{c} := \{ p \in M| r(p) = r_c \}$ is related
		to the extrinsic curvature of $\Sigma_c$, so we will
		define a normal covector to calculate the extrinsic curvature first.
		
		The normal covector is defined as

	\begin{equation}
	N_{a}=\frac{1}{\sqrt{g^{rr}}}(dr)_{a},
	\end{equation}
	where
	\begin{equation}
	\frac{1}{\sqrt{g^{rr}}}=\frac{1}{\sqrt{2\hat{\epsilon}}}r^{-\frac{1}{2}}%
	-\frac{(\hat{R}_{nlnl}+3\hat{\pi}_{I}^{2})}{4\hat{\epsilon}\sqrt
		{2\hat{\epsilon}}}r^{\frac{1}{2}}+O(r^{\frac{3}{2}}).
	\end{equation}
	The normal vector is
	\begin{equation}
	N^{a}=\frac{1}{\sqrt{g^{rr}}}(\partial_{t})^{a}+\sqrt{g^{rr}}(\partial
	_{r})^{a}+\frac{g^{ri}}{\sqrt{g^{rr}}}(\partial_{i})^{a}.
	\end{equation}
	With the normal vector, we can calculate the induced metric
	\begin{equation}
	h^{ab}=g^{ab}-N^{a}N^{b},\quad h_{b}^{a}=g^{ac}h_{cb},
	\end{equation}
	with
	\begin{subequations}
		\begin{align}
		h_{t}^{t} &  =1,\\
		h_{r}^{t} &  =-\frac{1}{2\hat{\epsilon}r}+\frac{(\hat{R}_{nlnl} + 3|\hat{\pi}|^{2})}{4\hat{\epsilon}^{2}}+O(r),\\
		h_{r}^{i} &  =\frac{\hat{\pi}_{I}\hat{e}_{I}^{i}}{\hat{\epsilon}}-\frac
		{1}{2\hat{\epsilon}}\left[  \hat{R}_{nInl}\hat{e}_{I}^{i}+3\hat{\theta}%
		_{IJ}^{\prime}\hat{\pi}_{I}\hat{e}_{J}^{i}+\frac{\hat{\pi}_{K}\hat{e}_{K}^{i}%
		}{\hat{\epsilon}}(\hat{R}_{nlnl}+3|\hat{\pi}|^{2})\right]  r+O(r^{2}),\\
		h_{j}^{i} &  =\delta_{j}^{i}.
		\end{align}
		The extrinsic curvature of $\Sigma_c$ is
	\end{subequations}
	\begin{equation}
	K_{ab}=\frac{1}{2}\mathcal{L}_{N}h_{ab}=-\frac{1}{\sqrt{g^{rr}}}\Gamma
	_{ab}^{r},
	\end{equation}
	where $a,b$ run through $t,x^{i}$ on $\Sigma_c$. Since $K_{ab}$ is the
	tensor on $\Sigma_c$, we use the induced metric $h^{ab}$ to
	raise its index to get,
	\begin{subequations}
		\label{extrinsic curvature}%
		\begin{align}
		K_{t}^{t}= &  \sqrt{\frac{\hat{\epsilon}}{2}}r^{-\frac{1}{2}}+\frac{\beta
		}{\sqrt{2\hat{\epsilon}}}r^{\frac{1}{2}}+O(r^{\frac{3}{2}}),\\
		K_{i}^{t}= &  -\frac{\hat{\pi}_{I}\hat{e}_{I}^{j}\hat{g}_{ij}}{\sqrt
			{2\hat{\epsilon}}}r^{-\frac{1}{2}}+\frac{\hat{\psi}_{i}}{(2\hat{\epsilon
			})^{\frac{3}{2}}}r^{\frac{1}{2}}+O(r^{\frac{3}{2}}),\\
		K_{j}^{i}= &  \frac{\xi_{j}^{i}}{\sqrt{2\hat{\epsilon}}}r^{\frac{1}{2}%
		}+O(r^{\frac{3}{2}}),\\
		K= &  \sqrt{\frac{\hat{\epsilon}}{2}}r^{-\frac{1}{2}}+\frac{(\beta+\xi)}%
		{\sqrt{2\hat{\epsilon}}}r^{\frac{1}{2}}+O(r^{\frac{3}{2}}),
		\end{align}
	\end{subequations}
	where
	\begin{subequations}
		\label{beta}%
		\begin{align}
		\beta= &  \frac{1}{4}(3\hat{R}_{nlnl}+|\hat{\pi}|^{2}),\\
		\hat{\psi}_{i}= &  -\left\{  \frac{1}{2}\tilde{\nabla}_{i}(\hat{R}%
		_{nlnl}-|\hat{\pi}|^{2})+2\hat{\pi}^{j}\left[  \tilde{\nabla}_{i}\hat{\pi
		}_j-\tilde{\nabla}_{j}\hat{\pi}_{i}\right]  \right. \nonumber\\
		&  -\left.  2\hat{\epsilon}(\hat{R}_{nInl}\hat{e}_{I}^{j}\hat{g}_{ij}%
		-\hat{\theta}_{IJ}^{\prime}\hat{\pi}_{I}\hat{e}_{J}^{j}\hat{g}_{ij})-\frac
		{1}{2}\hat{\pi}_{i}(\hat{R}_{nlnl}+3|\hat{\pi}|^{2})\right\}  ,\\
		\xi_{j}^{i}= &  -2\hat{g}^{ik}\tilde{\nabla}_{(j}\hat{\pi}_{k)}+2\hat{\pi}%
		^{i}\hat{\pi}_{j}+2\hat{\epsilon}\hat{\theta}_{IJ}^{\prime}\hat{e}_{I}^{i}%
		\hat{e}_{J}^{m}\hat{g}_{jm},\\
		\xi= &  h^{ij}\xi_{ij},
		\end{align}
		and
	\end{subequations}
	\begin{equation}
	\hat{\pi}_{i}=\hat{\pi}_{I}\hat{e}_{I}^{j}\hat{g}_{ij},\quad\hat{\pi}^{i}%
	=\hat{\pi}_{I}\hat{e}_{I}^{i}.
	\end{equation}
	Note that the derivative with respect to $t$ vanishes on $\mathcal{H}$.
	
	\subsection{Brown-York Tensor in Near Horizon and Non-Relativistic Limit}

	According to the method proposed by \cite{1104.5502}, we need to apply the Petrov-like condition $C_{\mu\nu\rho\sigma} l^{\mu} E^{\nu}_{i} l^{\rho} E^{\sigma}_{j}|_{\Sigma_{c}} = 0$ on the hypersurface and take the near horizon and non-relativistic limit of $\Sigma_{c}$ to get the Navier-Stokes equation.
	To do so, we introduce a rescaling parameter $\lambda$ and define a new time coordinate $\tau= 2\hat{\epsilon}\lambda^{2} t$.
	We also choose the radius of $\Sigma_c$ to be $r_{c} = 2\hat{\epsilon}\lambda^{2}$.
	
	The Brown-York tensor is
	\begin{equation}
	t_{b}^{a(B)}=Kh_{b}^{a}-K_{b}^{a},
	\end{equation}
    where $t_{b}^{a(B)}$ is the Brown-York tensor of the background. After fixing $r=r_{c}$ and changing $(t,x^{i})$
	to $(\tau,x^{i})$, $t_{b}^{a(B)}$ can be expressed in $\lambda$ as,
	\begin{subequations}
		\begin{align}
		t_{\tau}^{\tau(B)}= &  \xi\lambda+O(\lambda^{3}),\\
		t_{i}^{\tau(B)}= &  \hat{\pi}_i\lambda
		+O(\lambda^{3}),\\
		t_{\tau}^{i(B)}= &  O(\lambda),\\
		t_{j}^{i(B)}= &  \frac{1}{2\lambda}\delta_{j}^{i}+\left[  (\beta+\xi
		)\delta_{j}^{i}-\xi_{j}^{i}\right]  \lambda+O(\lambda^{3}),\\
		t^{(B)}= &  \frac{p}{2\lambda}+p(\beta+\xi)\lambda+O(\lambda^{3}).
		\end{align}
	\end{subequations}
	Adding perturbations to Brown-York tensor,
	\begin{equation}
	t_{b}^{a}=t_{b}^{a(B)}+\sum_{k=1}t_{b}^{a(k)}\lambda^{k},
	\end{equation}
we obtain
	\begin{subequations}
		\begin{align}
		t_{\tau}^{\tau}= &  \left[  \xi+t_{\tau}^{\tau(1)}\right]  \lambda
		+O(\lambda^{2}),\\
		t_{i}^{\tau}= &  \left[  \hat{\pi}_i + t_{i}^{\tau(1)} \right] \lambda + O(\lambda^{2}),\\
		t_{\tau}^{i}= &  -\hat{g}^{ij}t_{j}^{\tau(1)}\lambda^{-1}+O(\lambda),\\
		t_{j}^{i}= &  \frac{1}{2\lambda}\delta_{j}^{i}+\left[  (\beta+\xi)\delta
		_{j}^{i}-\xi_{j}^{i}+t_{j}^{i(1)}\right]  \lambda+O(\lambda^{2}),\\
		t= &  \frac{p}{2\lambda}+\left[  p(\beta+\xi)+t^{(1)}\right]  \lambda
		+O(\lambda^{2}),\\
		t_{\tau j}= &  \left[  \hat{\pi}_j + t_{j}^{\tau(1)} \right] \lambda^{-1} + O(\lambda),\\
		t_{\tau\tau}= &  \left[  -(\xi+t_{\tau}^{\tau(1)})-2\hat{\pi}^mt_m^{\tau(1)}\right]  \lambda^{-1}+O(\lambda),\\
		t_{ij}= &  \frac{1}{2}\hat{g}_{ij}\lambda^{-1}+O(\lambda).
		\end{align}
		We also add the perturbation terms to $\phi$ and $A_{a}$ as for Brown-York tensor:
	\end{subequations}
	\begin{subequations}
		\begin{align}
		\phi =&  \phi_{(0)} + \phi^{P(1)}\lambda + \left[ 2\hat{\epsilon}\phi_{(1)} + \phi^{P(2)}\right] \lambda^{2} + O(\lambda^{3}),\\
		A_{a} =& A_{a(0)} + A_{a}^{P(1)} \lambda + \left[ 2\hat{\epsilon}A_{a(1)} + A_{a}^{P(2)} \right] \lambda^{2} + O(\lambda^{3}),
		\end{align}
		and
	\end{subequations}
	\begin{align}
	J_{ab} =& J_{ab(0)} + J_{ab}^{P(1)}\lambda + \left[ J_{ab(1)} + J_{ab}^{P(2)} \right] \lambda^{2} + O(\lambda^{3}), nn\\
	\Lambda^{\prime} =& \Lambda_{(0)}^{\prime} + \Lambda^{\prime P(1)} \lambda + O(\lambda^{2}),
	\end{align}
	where $J_{ab}^{P(k)}$ and $\Lambda'_{(0)}$ are perturbation terms. Here we only perturb $\phi$ field and $A_{a}$ field with the metric unperturbed.
	Then using Eq.(\ref{Ricci0}), we have the components of Ricci tensor
	\begin{subequations}
		\begin{align}
		R_{\tau\tau}= &  -\Lambda_{(0)}^{\prime}\lambda^{-2}-\Lambda^{\prime
			P(1)}\lambda^{-1}+O(1),\\
		R_{\tau r}= &  \frac{\Lambda_{(0)}^{\prime}}{2\hat{\epsilon}}\lambda
		^{-2}+\frac{\Lambda^{\prime P(1)}}{2\hat{\epsilon}}\lambda^{-1}+O(1),\\
		R_{\tau i}= &  \left[  2\Lambda_{(0)}^{\prime}\hat{\pi}_{i}+J_{\tau
			i(0)}\right]  +\left[  2\Lambda^{\prime P(1)}\hat{\pi}_{i}+J_{\tau i}%
		^{P(1)}\right]  \lambda+O(\lambda^{2}),\\
		R_{ij}= &  \left[  \Lambda_{(0)}^{\prime}\hat{g}_{ij}+J_{ij(0)}\right]
		+\left[  \Lambda^{\prime P(1)}\hat{g}_{ij}+J_{ij}^{P(1)}\right]
		\lambda+O(\lambda^{2}),\\
		R_{rr}= &  J_{rr(0)}+J_{rr}^{P(1)}\lambda+O(\lambda^{2}),\\
		R_{ri}= &  J_{ri(0)}+J_{ri}^{P(1)}\lambda+O(\lambda^{2}),\\
		R= &  R_{(0)}+O(\lambda),
		\end{align}
	\end{subequations}
	where
	\begin{equation}
	R_{(0)} = \frac{2(p+2)}{p}\Lambda + \frac{(p+2)}{2p}(V)_{(0)} + \frac{1}{2}(|D\phi|^2)_{(0)} + \frac{(p-2)}{4p}(fF^2)_{(0)}.
	\end{equation}

	\subsection{Petrov-like Condition}
	
	\label{petrov-like condition} Now, we are going to calculate the Petrov-like
	condition on the hypersurface $\Sigma_{c}$. The Petrov-like condition is
	\begin{equation}
	l^{\mu}E_{i}^{\nu}l^{\rho}E_{j}^{\sigma}C_{\mu\nu\rho\sigma}|_{\Sigma_{c}%
	}=C_{lilj}|_{\Sigma_{c}}=0.
	\end{equation}
	From the tetrad and normal vector defined in the last section, we have
	\begin{equation}
	l^{a}=\frac{U}{\sqrt{g^{rr}}}N^{a}+2\hat{\epsilon}\lambda^{2}\left(
	1-\frac{U}{g^{rr}}\right)  (\partial_{\tau})^{a}+\left(  X^{i}-\frac{Ug^{ri}%
	}{g^{rr}}\right)  (\partial_{i})^{a},
	\end{equation}
	and the Petrov-like condition becomes
	\begin{align}
	\label{petrov}
	0= &  \left[  \frac{U^{2}}{g^{rr}}C_{NiNj}+\left(  1-\frac{U}{g^{rr}}\right)
	^{2}C_{\tau i\tau j}+\frac{U}{\sqrt{g^{rr}}}\left(  1-\frac{U}{g^{rr}}\right)
	(C_{Ni\tau j}+C_{Nj\tau i})\right. \nonumber\\
	&  +\frac{U}{\sqrt{g^{rr}}}\left(  X^{k}-\frac{Ug^{rk}}{g^{rr}}\right)
	(C_{Nikj}+C_{Njki})\nonumber\\
	&  +\left(  X^{k}-\frac{Ug^{rk}}{g^{rr}}\right)  \left(  X^{m}-\frac{Ug^{rm}%
	}{g^{rr}}\right)  C_{kimj}\nonumber\\
	&  +\left.  \left(  1-\frac{U}{g^{rr}}\right)  \left(  X^{k}-\frac{Ug^{rk}%
	}{g^{rr}}\right)  (C_{\tau ikj}+C_{\tau jki})\right]  _{\Sigma_{c}}.
	\end{align}
	Note that Riemann tensor can be expressed by Weyl tensor and Ricci tensor as,
	\begin{equation}
	R_{abcd}=C_{abcd}+\frac{1}{p}(g_{ac}R_{bd}+g_{bd}R_{ac}-g_{ad}R_{bc}%
	-g_{bc}R_{ad})-\frac{R}{p(p+1)}(g_{ac}g_{bd}-g_{ad}g_{bc}).
	\end{equation}
	From Gauss's equation, we know the Riemann curvature of the $(p+1)$-dimensional
	hypersurface embedded in the $(p+2)$-dimensional space-time can be expressed in terms
	of the Riemann curvature and the extrinsic curvature of the $(p+2)$-dimensional
	space-time as follows,
	\begin{subequations}
		\begin{align}
		h_{a}^{\alpha}h_{b}^{\beta}h_{c}^{\gamma}h_{d}^{\delta}R_{\alpha\beta
			\gamma\delta} &  =\bar{R}_{abcd}-K_{ac}K_{bd}+K_{ad}K_{bc},\\
		h_{a}^{\alpha}h_{b}^{\beta}h_{c}^{\gamma}N^{\delta}R_{\alpha\beta\gamma\delta}
		&  =\bnabla_aK_{bc}-\bnabla_bK_{ac},\\
		h_{a}^{\alpha}N^{\beta}h_{b}^{\gamma}N^{\delta}R_{\alpha\beta\gamma\delta} &
		=-\bar{R}_{ab}+KK_{ab}-K_{ac}K_{b}^{~c},
		\end{align}
		where greek indices denote the coordinate $(\tau, r, x^i)$ in $(p+2)$-dimensional space-time and
		latin indices denote the coordinate $(\tau, x^i)$ on the $(p+1)$-dimensional hypersurface $\Sigma_{c}$.
		``$\bnabla$" is the covariant derivative on $(p+1)$-dimensional hypersurface $\Sigma_c$.
		\textquotedblleft$\bar{R}_{abcd}$" and \textquotedblleft$\bar{R}_{ab}$" are the Riemann tensor and the Ricci tensor of $\Sigma_{c}$ which can be calculated by induced metric $h_{ab}$. From these two equation we have
	\end{subequations}
	\begin{subequations}
	\label{weyl}
		\begin{align}
		h_{a}^{\alpha}h_{b}^{\beta}h_{c}^{\gamma}h_{d}^{\delta}C_{\alpha\beta
			\gamma\delta}= &  \bar{R}_{abcd}-K_{ac}K_{bd}+K_{ad}K_{bc}\nonumber\\
		&  -\left[  \frac{1}{p}(h_{ac}h_{b}^{\alpha}h_{d}^{\beta}R_{\alpha\beta
		}+h_{bd}h_{a}^{\alpha}h_{c}^{\beta}R_{\alpha\beta}-h_{ad}h_{b}^{\alpha}%
		h_{c}^{\beta}R_{\alpha\beta}-h_{bc}h_{a}^{\alpha}h_{d}^{\beta}R_{\alpha\beta
		})\right. \nonumber\\
		&  -\left.  \frac{R}{p(p+1)}(h_{ac}h_{bd}-h_{ad}h_{bc})\right]  ,\\
		h_{a}^{\alpha}h_{b}^{\beta}h_{c}^{\gamma}N^{\delta}C_{\alpha\beta\gamma\delta
		}= &  \bnabla_{a}K_{bc}-\bnabla_{b}K_{ac}\nonumber\\
		&  -h_{a}^{\alpha}h_{b}^{\beta}h_{c}^{\gamma}N^{\delta}\left[  \frac{1}%
		{p}(g_{\alpha\gamma}R_{\beta\delta}+g_{\beta\delta}R_{\alpha\gamma}%
		-g_{\alpha\delta}R_{\beta\gamma}-g_{\beta\gamma}R_{\alpha\delta})\right.
		\nonumber\\
		&  -\left.  \frac{R}{p(p+1)}(g_{\alpha\gamma}g_{\beta\delta}-g_{\alpha\delta
		}g_{\beta\gamma})\right]  ,\\
		h_{a}^{\alpha}N^{\beta}h_{b}^{\gamma}N^{\delta}C_{\alpha\beta\gamma\delta}= &
		-\bar{R}_{ab}+KK_{ab}-K_{ac}K_{b}^{c}+h_{a}^{\alpha}h_{b}^{\gamma}%
		R_{\alpha\gamma}\nonumber\\
		&  -h_{a}^{\alpha}h_{b}^{\gamma}N^{\beta}N^{\delta}\left[  \frac{1}%
		{p}(g_{\alpha\gamma}R_{\beta\delta}+g_{\beta\delta}R_{\alpha\gamma}%
		-g_{\alpha\delta}R_{\beta\gamma}-g_{\beta\gamma}R_{\alpha\delta})\right.
		\nonumber\\
		&  -\left.  \frac{R}{p(p+1)}(g_{\alpha\gamma}g_{\beta\delta}-g_{\alpha\delta
		}g_{\beta\gamma})\right]  .
		\end{align}
		\end{subequations}
		Substituting the Weyl tensor (\ref{weyl}) into Petrov-like condition (\ref{petrov}) and using
	
	\begin{equation}
	K_{b}^{a}=\frac{t}{p}h_{b}^{a}-t_{b}^{a}%
	\end{equation}
	to rewrite the extrinsic curvature in terms of Brown-York tensor, we obtain
	\begin{align}
	0= &  \frac{U^{2}}{g^{rr}}\left\{  \frac{t}{p}t_{j}^{i}-t_{c}^{i}t_{j}%
	^{c}-\frac{t}{p}h^{i\tau}t_{\tau j}+h^{i\tau}t_{\tau c}t_{j}^{c}-\bar{R}%
	_{qj}h^{iq}+h^{iq}h_{q}^{\alpha}h_{j}^{\gamma}R_{\alpha\gamma}\right.
	\nonumber\\
	&  -\left.  h^{iq}h_{q}^{\alpha}h_{j}^{\gamma}N^{\beta}N^{\delta}\left[
	\frac{1}{p}(g_{\alpha\gamma}R_{\beta\delta}+g_{\beta\delta}R_{\alpha\gamma
	}-g_{\alpha\delta}R_{\beta\gamma}-g_{\beta\gamma}R_{\alpha\delta})-\frac
	{R}{p(p+1)}(g_{\alpha\gamma}g_{\beta\delta}-g_{\alpha\delta}g_{\beta\gamma
	})\right]  \right\} \nonumber\\
	&  +4\hat{\epsilon}^{2}\lambda^{4}\left(  1-\frac{U}{g^{rr}}\right)
	^{2}\left\{  -h_{\tau\tau}(\frac{t^{2}}{p^{2}}h_{j}^{i}-\frac{t}{p}t_{j}%
	^{i})+\frac{t}{p}h_{j}^{i}t_{\tau\tau}-t_{\tau\tau}t_{j}^{i}+t_{\tau j}%
	t_{\tau}^{i}+\bar{R}_{\tau q\tau j}h^{iq}\right. \nonumber\\
	&  -\left[  \frac{1}{p}h^{iq}\left(  h_{\tau\tau}h_{j}^{\alpha}h_{q}^{\beta
	}R_{\alpha\beta}+h_{jq}h_{\tau}^{\alpha}h_{\tau}^{\beta}R_{\alpha\beta
}-h_{\tau j}h_{\tau}^{\alpha}h_{q}^{\beta}R_{\alpha\beta}-h_{\tau q}h_{\tau
}^{\alpha}h_{j}^{\beta}R_{\alpha\beta}\right)  \right. \nonumber\\
&  -\left.  \left.  \frac{R}{p(p+1)}h^{iq}(h_{\tau\tau}h_{jq}-h_{\tau
	j}h_{\tau q})\right]  \right\} \nonumber\\
&  +2\hat{\epsilon}\lambda^{2}\frac{U}{\sqrt{g^{rr}}}\left(  1-\frac{U}%
{g^{rr}}\right)  \left\{  h^{iq}\left[  h_{\tau q}D_{j}\frac{t}{p}+h_{\tau
	j}\bnabla_{q}\frac{t}{p}-2h_{jq}\bnabla_{\tau}\frac{t}{p}-\bnabla_{j}t_{\tau q}-\bnabla_{q}t_{\tau
	j}+2\bnabla_{\tau}t_{jq}\right]  \right. \nonumber\\
&  -h^{iq}h_{j}^{\alpha}h_{\tau}^{\beta}h_{q}^{\gamma}N^{\delta}\left[
\frac{1}{p}(g_{\alpha\gamma}R_{\beta\delta}+g_{\beta\delta}R_{\alpha\gamma
}-g_{\alpha\delta}R_{\beta\gamma}-g_{\beta\gamma}R_{\alpha\delta})-\frac
{R}{p(p+1)}(g_{\alpha\gamma}g_{\beta\delta}-g_{\alpha\delta}g_{\beta\gamma
})\right] \nonumber\\
&  -\left.  h^{iq}h_{q}^{\alpha}h_{\tau}^{\beta}h_{j}^{\gamma}N^{\delta}
\left[  \frac{1}{p}(g_{\alpha\gamma}R_{\beta\delta}+g_{\beta\delta}%
R_{\alpha\gamma}-g_{\alpha\delta}R_{\beta\gamma}-g_{\beta\gamma}%
R_{\alpha\delta})-\frac{R}{p(p+1)}(g_{\alpha\gamma}g_{\beta\delta}%
-g_{\alpha\delta}g_{\beta\gamma})\right]  \right\}  .
\end{align}
Focusing on the first nonzero order $O(\lambda
^{2})$ and neglecting the higher order terms in $\lambda$, we have
\begin{subequations}
	\begin{align}
	\frac{U}{\sqrt{g^{rr}}}= &  \hat{\epsilon}\lambda+\frac{\hat{\epsilon}}%
	{2}\left( \hat{R}_{nlnl} + |\hat{\pi}|^{2}\right)  \lambda^{3}+O(\lambda
	^{5}),\\
	1-\frac{U}{g^{rr}}= &  \frac{1}{2}+\frac{1}{2}|\hat{\pi}|^2 \lambda
	^{2}+O(\lambda^{4}),\\
	X^{k}-\frac{Ug^{rk}}{g^{rr}}= &  2\hat{\epsilon}^{2}\left(  \hat{\theta}%
	_{IJ}^{\prime}\hat{\pi}_{I}\hat{e}_{J}^{k}-\frac{|\hat{\pi}|^{2}}{\hat{\epsilon}} \hat{\pi}^k \right)  \lambda^{4} + O(\lambda^{6}).
	\end{align}
	After some strait-forward calculations, the $O(\lambda^{2})$ terms of Petrov-like condition is obtained,
\end{subequations}
\begin{align} \label{Petrov}
t_{j}^{i(1)}= &  -2\hat{g}^{ik}\tilde{\nabla}_{(j}t_{k)}^{\tau(1)}+2\hat
{g}^{im}t_{m}^{\tau(1)}\left(  \hat{\pi}_{j}+t_{j}^{\tau(1)}\right)  -2\hat
{g}^{ik}\tilde{\nabla}_{(j}\hat{\pi}_{k)}+\frac{t^{(1)}}{p}\delta_{j}%
^{i}\nonumber\\
&  +\xi_{j}^{i}-\tilde{R}_{j}^{i}+\left[  \Lambda_{(0)}^{\prime}\delta_{j}%
^{i}+\hat{g}^{iq}J_{jq(0)}\right]  .
\end{align}
Here $\tnabla$ is the covariant derivative on $p$-dimensional section $S^p$ and $\tilde{R}_{ij}$ is the Ricci tensor on the $p$-dimensional sphere $S^p$ in $(p+1)$-dimensional hypersurface $\Sigma_{c}$ which can be calculated from the metric $\hat{g}_{ij}$, $i, j = 1 \cdots p$. Now we see that the higher order terms of $\phi$ field and $A_a$ field do not contribute any effect.

%	The components of tetrad are:
%	\begin{subequations}
%	\begin{align}
%		R_{nn} =& n^a n^b R_{ab} = R_{rr}, \\
%		R_{nl} =& n^a l^b R_{ab} = R_{rt} + UR_{rr} + 	X^iR_{ri}, \\
%		R_{nI} =& n^a E_I^b R_{ab} = W_IR_{rr} + e^i_IR_{ri}.
%	\end{align}
%	\end{subequations}
%	and the components of Riemann tensor can be obtained by:
%	\begin{equation}
%		R_{abcd} = C_{abcd} + \frac{1}{p}\Big[g_{ac}R_{bd} + g_{bd}R_{ac} - g_{ad}R_{bc} - g_{bc}R_{ad}\Big] - \frac{R}{p(p+1)}\Big[g_{ac}g_{bd} - g_{ad}g_{bc}\Big].
%	\end{equation}
%	Therefore we have
%	\begin{equation} \label{Riemann}
%	\begin{split}
%		R_{nlnl} =& C_{nlnl} - \frac{2}{p}R_{nl} + \frac{R}{p(p+1)}, \\
%		R_{nInl} =& C_{nInl} - \frac{1}{p}R_{nI}, \\
%		R_{nInJ} =& C_{nInJ} + \frac{1}{p}\delta_{IJ}R_{nn}.
%	\end{split}
%	\end{equation}
%

%TCIMACRO{\TeXButton{equation number}{\setcounter{equation}{0}
%\renewcommand{\theequation}{\arabic{section}.\arabic{equation}}}}%
%BeginExpansion
\setcounter{equation}{0}
\renewcommand{\theequation}{\arabic{section}.\arabic{equation}}%
%EndExpansion

\section{From Petrov-like Condition to Navier Stokes Equation}
\label{IV}

To get the Navier-Stokes equation, we will substitute Eq.(\ref{Petrov}) into Gauss-Codazzi equation. First, we calculate the Gauss-Codazzi equation,
\begin{equation}
\bar{\nabla}_{a}t^{a}_{b} = 0.
\end{equation}
Writing the components explicitly we have:
\begin{equation}
\bar{\nabla}_{\tau}t^{\tau}_{\tau}+ \bar{\nabla}_{i}t^{i}_{\tau}=0.
\end{equation}
The equation in order ($1/\lambda) $ reads
\begin{equation}
\label{incompressible}0 = -\tilde{\nabla}_{i}(\hat{g}^{ij} t^{\tau(1)}_{j}),
\end{equation}
which tells us that, if we identify $t^{\tau(1)}_{j}$ to be the velocity $v_{j}$, the fluid is incompressible.
The other components are:
\begin{equation}
\bar{\nabla}_{\tau}t^{\tau}_{i} + \bar{\nabla}_{j}t^{j}_{i} =0.
\end{equation}
The equation in order $\lambda$ reads,
\begin{align}
\label{Codazzi}0 = &  \partial_{\tau}t^{\tau(1)}_{i} - 2 t^{\tau j (1)}%
\tilde{\nabla}_{(i}\hat{\pi}_{j)} - 2 t^{\tau j (1)}\tilde{\nabla}_{[i}%
\hat{\pi}_{j]}\nonumber\\
& - \tilde{\nabla}_{j}\xi^{j}_{i} + \tilde{\nabla}_{i}(\beta+ \xi) - \frac
{1}{4}\tilde{\nabla}_{i}(\hat{R}_{nlnl} - |\hat{\pi}|^{2})\nonumber\\
& + \tilde{\nabla}_{j} t^{j (1)}_{i},
\end{align}
From Gaussian equation,
\begin{equation}
\bar{R}+K^{ab}K_{ab}-K^{2}=0,
\end{equation}
the zeroth order of $\lambda$ gives us,
\begin{equation}
t_{\tau}^{\tau(1)} = -2\hat{g}^{ij}t_{i}^{\tau(1)}t_{j}^{\tau(1)} - 2\hat{\pi}^it_{i}^{\tau(1)} - \xi + \tilde{R}.
\end{equation}
It's the Hamiltonian constraint in the energy-momentum for the corresponding
fluid. Thus it is natural to consider the corresponding as,
\begin{equation}
t_{i}^{\tau(1)}\longleftrightarrow\frac{1}{2}v_{i}\text{, \ \ \ \ }%
\frac{t^{(1)}}{p}\longleftrightarrow\frac{P}{2}.
\end{equation}
Substituting equation (\ref{Petrov}) into equation (\ref{Codazzi}) and by
the correspondence, we have
\begin{align}
0= &  \frac{1}{2}\partial_{\tau}v_{i}+\frac{1}{2}\tilde{\nabla}_{i}P+\frac
{1}{2}v^{j}\tilde{\nabla}_{j}v_{i}-\frac{1}{2}\tilde{\nabla}^{k}\tilde{\nabla
}_{k}v_{i}-\frac{1}{2}\tilde{R}_{im}v^{m}-2v^{j}\tilde{\nabla}_{[i}\hat{\pi
}_{j]}-2\tilde{\nabla}^{k}\tilde{\nabla}_{(i}\hat{\pi}_{k)}\nonumber\\
&  -\tilde{\nabla}_{j}\tilde{R}_{i}^{j}+\tilde{\nabla}_{i}(\beta+\xi)-\frac
{1}{4}\tilde{\nabla}_{i}(\hat{R}_{nlnl}-|\hat{\pi}|^2)\nonumber\\
&  +\left[  \tilde{\nabla}_{i}\Lambda_{(0)}^{\prime}+\hat{g}^{jq}\tilde
{\nabla}_{j}J_{iq(0)}\right]  .
\end{align}
From the constraint equation of spin-connections of
IH derived in \cite{0111067}, $2\hat{\epsilon}\hat{\theta}_{IJ}^{\prime}$ can
be written as,
\begin{equation}
2\hat{\epsilon}\hat{\theta}_{II}^{\prime} = -\tilde{\nabla}_{k}\hat{\pi}%
^{k} + 2|\hat{\pi}|^2,
\end{equation}
and
\begin{equation}
\hat{R}_{nlnl}=\hat{C}_{nlnl}-\frac{2}{p}\hat{R}_{nl}+\frac{\hat{R}}%
{p(p+1)}=\hat{C}_{nlnl}-\frac{2}{p}\Lambda_{(0)}^{\prime}+\frac{R_{(0)}%
}{p(p+1)}.
\end{equation}
Using Eq.(\ref{beta}), we have
\begin{align}
\label{NVE}
0 =& \frac{1}{2}\partial_{\tau}v_{i} + \frac{1}{2}\tilde{\nabla}_{i}P + \frac{1}{2}v^{j}\tilde{\nabla}_{j}v_{i} - \frac{1}{2}\tilde{\nabla}^{k}\tilde{\nabla
}_{k}v_{i} - \frac{1}{2}\tilde{R}_{im}v^{m} - 2v^{j}\tilde{\nabla}_{[i}\hat{\pi}_{j]} \nonumber\\
&-2\tilde{\nabla}^{k}\tilde{\nabla}_{k}\hat{\pi}_{i} - 2\tilde{R}_{im}\hat{\pi}^{m} - \tilde{\nabla}_{j}\tilde{R}_{i}^{j} + \frac{1}{2}\tilde{\nabla}_{i} \left( \hat{C}_{nlnl} + 9|\hat{\pi}|^{2} - 8\tilde{\nabla}_{j}\hat{\pi}^{j} \right) \nonumber\\
&+ \left[ \frac{(p-1)}{p} \tilde{\nabla}_{i}\Lambda_{(0)}^{\prime} + \hat{g}^{jq} \tilde{\nabla}_{j}J_{iq(0)} + \frac{1}{2p(p+1)} \tilde{\nabla}_{i}R_{(0)} \right].
\end{align}
This is the Navier-Stokes equation of incompressible fluid with rotation and external force terms.
The last term in the first line is the Coriolis force term.
The terms in the third line of equation (\ref{NVE}) are
external force terms caused by the scalar field and electromagnetic field.

To be clearer, we consider the case that the $p$-dimensional geometry described by $\hat{g}_{ij}$ is flat, which implies that $\tilde{R}_{ij}=0$. Furthermore, if we take
$p=3$ and define the angular velocity of the reference frame as,
\begin{equation}
\mathbf{\Omega} = \tilde{\nabla} \times \mathbf{\hpi},
\end{equation}
the Eq.(\ref{NVE}) along with Eq.(\ref{incompressible}) becomes
\begin{subequations}
	\begin{align}
	0= &  \tilde{\nabla}\cdot\mathbf{v},\\
	0= &  \partial_{\tau}\mathbf{v} + \mathbf{v}\cdot\tilde{\nabla}\mathbf{v} + \tilde{\nabla}P - \tilde{\nabla}^{2}\mathbf{v} + 2\mathbf{\Omega}\times \mathbf{v} + \mathbf{f},
	\end{align}
\end{subequations}
with the external force term being
\begin{align}
\label{NVE2}
\mathbf{f}_i =& -4\tilde {\nabla}^{2} \mathbf{\hat{\pi}}_i - 8\tilde{\nabla}_i%
(\tilde{\nabla}\cdot\mathbf{\hat{\pi}})+\tilde{\nabla}_i(\hat{C}_{nlnl} + 9|\mathbf{\hat{\pi}}|^{2}) \nonumber\\
&+ \frac{1}{48}\tnabla_i |D\phi|^2_{(0)} + \frac{1}{2}\tnabla^j \Big[ (D_i\phi)^*(D_j\phi) \Big]_{(0)} + \frac{7}{48}\tnabla_iV(\phi, \phi^*)_{(0)}  \nn\\
&- \frac{5}{96}\tnabla_i \Big( f(\phi, \phi^*)F^2 \Big)_{(0)} + \frac{1}{2}\tnabla_j \Big( f(\phi, \phi^*)F_{ic}F^{jc} \Big)_{(0)}.
\end{align}
We note that, to calculate the first two terms in the second line of Eq.(\ref{NVE2}) we have to know the gauge field $A_{a(0)}$. The true physical initial data on the horizon is $F_{ab(0)}$. Similar to the analysis in \cite{1311.3384}, $A_{a(0)}$ can be fixed from $F_{ab(0)}$ by choosing a proper gauge.

\section{Conclusion and Discussion}
\label{V}

In this paper, we showed that the effects on the dual fluid by adding the scalar and electromagnetic fields into a space-time background are due to non-zero Ricci tensors determined by Einstein equation.
If we turn off $\phi$, $F_{ab}$ and $\pi_I$, which means we restrict our case to a non-rotating black hole in a vacuum space-time, Eq.(\ref{NVE}) reduces to the standard Navier-Stokes equation which agrees with the result in \cite{1303.3736}.
If we only consider the rotation of black hole, our result agrees with the Navier-Stokes equation in \cite{1511.08691}.

In \cite{1401.6487}, the authors considered a spherical symmetric metric with a $\phi$ field, which becomes a constant on the horizon, and concluded that the scalar field does not contribute to the external force in the dual fluid.
However in our case, we do not demand the spherical symmetry of the space-time and the scalar field is not isometric on the horizon. Therefore, the scalar field indeed implies the external force in Eq.(\ref{NVE2})
	
On the other hand, if we only consider the electromagnetic field and set $f(\phi, \phi^*) = 1$, the form of the external force terms we obtained agrees with that of the magnetohydrodynamics equations in \cite{1310.4181}.

There are still some interesting problems need to be considered. From Eq. (\ref{NVE}), we can see that the perturbation terms of the matter fields have no influence on the Navier-Stokes equation in $O(\lambda^{2})$ terms of Gauss-Codazzi equation. If we consider higher order terms, maybe more non-linear effects can be seen. Until now, all dual fluid equation on horizon are incompressible, so one can not see the bulk viscosity effect of the dual fluid. Is it possible to get a hydrodynamics equation of compressible fluid if we choose some other kinds of boundary conditions ? We hope to address these questions in our future research.

\begin{acknowledgments}
	This work is partially supported by the
	National Science Council (NSC 101-2112-M-009-005) and National Center for
	Theoretical Science, Taiwan. X. Wu is supported by the National Natural
	Science Foundation of China (Grant Nos. 11475179, 11175245 and 11575286).
\end{acknowledgments}

%\bibliographystyle{unsrt}
%\bibliography{FGCbibtex}

\end{document}